
\magnification=\magstep1
\baselineskip=20 truept
\def\ni{\noindent}
\def\bn{\bigskip\noindent}
\def\sn{\smallskip\noindent}
\def\pa{\partial}
\def\half{{\textstyle{1\over2}}}
\def\vp{\varphi}
\nopagenumbers
\font\scap=cmcsc10

\rightline{DTP/95/1; NI94032}
\vskip 1truein
\centerline{\bf Stable Topological Skyrmions on the 2D Lattice.}

\vskip 1truein
\centerline{\scap R. S. Ward}
\bn\centerline{\sl Isaac Newton Institute for Mathematical Sciences,}
\centerline{\sl Cambridge CB3 0EH, UK}
\sn\centerline{and}
\sn\centerline{\sl Dept of Mathematical Sciences, University of Durham,}
\centerline{\sl Durham DH1 3LE, UK.$^*$}

\vskip 2truein
\ni{\bf Abstract.} In the continuum O(3) sigma model in two spatial
dimensions, there are topological solitons whose size can be stabilized by
adding Skyrme and potential terms. This paper describes a lattice version,
namely a natural way of
modifying the 2d Heisenberg model to achieve topological stability on
the lattice.

\bn To appear in Letters in Mathematical Physics.
\vfill
\hrule
\sn$^*$ Permanent address.
\eject

\def\mut{\tilde\mu}

\pageno=2
\footline={\hss\tenrm\folio\hss}

\ni{\bf1. Introduction.}
\sn Topological solitons in sigma models, and specifically in the
two-dimensional O(3) sigma model, have been studied in several different
contexts:
\item{$\bullet$} as metastable states in two-dimensional ferromagnets [1,2];
\item{$\bullet$} as instantons in a field theory which models four-dimensional
                 gauge theory [3];
\item{$\bullet$} as a model for particles, in particular as a simpler version
                 of the three-dimensional skyrme model [4,5].

\ni In each of these applications, the lattice version of the model (where the
field is defined on a two-dimensional lattice), is relevant. For
condensed-matter applications, the relevance is direct; while for
field-theory applications, a discrete version is needed in order to do
numerical computations [6,7,8,9,4,5]. In the usual lattice version of the
O(3) sigma model, namely the Heisenberg model, the topological properties
are essentially lost (which is what one would expect, since topology has to
do with continuity). But in this paper we shall see how topology (and
topological stability of the solitons) can be maintained on the lattice.

\bn{\bf2. A Topological Heisenberg Model.}
\sn Let $L$ be an $n_x\times n_y$ lattice; in other words, there are $n_xn_y$
lattice sites, labelled by integers $(x,y)$ with $1\leq x\leq n_x$ and
$1\leq y \leq n_y$. The field $\vp$ is a map from $L$ to $S^2$. Think of
$\vp_{x,y}$, the field (or spin)
at the lattice site $(x,y)$, as a three-dimensional
unit vector. The energy of $\vp$ is defined by the expression
$$\eqalign{
  E  &= {1\over4\pi}\sum_{x=1}^{n_x-1}\sum_{y=1}^{n_y}
           f(\vp_{x,y}\cdot\vp_{x+1,y})
        +{1\over4\pi}\sum_{x=1}^{n_x}\sum_{y=1}^{n_y-1}
           f(\vp_{x,y}\cdot\vp_{x,y+1}) \cr
     &\qquad\qquad +{\mu^2\over4\pi}\sum_{x=1}^{n_x}\sum_{y=1}^{n_y}
           (1-\vp_{x,y}\cdot k). \cr
}\eqno(1)$$
Here $f$ is a real-valued function with $f(1)=0$ and $f'(1)=-1$,
the dot denotes the Euclidean scalar product, $\mu$ is a constant scalar,
and $k$ is a constant unit vector.

First, let us examine the continuum limit. To this end, let $n_x$ and $n_y$
be infinite, introduce a lattice spacing $h$ (so that $x$ and $y$ temporarily
become integer multiples of $h$), and replace $\mu$ by $\mut
=\mu/h$. Then the $h\to0$ limit of (1) is
$$E_{\rm cont} = {1\over4\pi}\int_{\bf R^2}
                 \bigl[\half(\pa_x\vp)^2 + \half(\pa_y\vp)^2
                       + \mut^2(1-\vp\cdot k)\bigr]\,dx\,dy.\eqno(2)$$
If $\mut=0$, this is the two-dimensional O(3) sigma model; the
$\mut$ term is an optional potential. Notice that the function $f$
does not appear in (2): any $f$ which is smooth near 1, with $f(1)=0$ and
$f'(1)=-1$, leads to (2).

The boundary condition is $\vp\to k$ as $r\to\infty$ on ${\bf R^2}$
(sufficiently fast for $E_{\rm cont}$ to converge). The configuration
space of smooth fields $\vp$ satisfying this boundary condition is
disconnected: its components are labelled by an integer $N$, the winding
number or topological charge of $\vp$. A version of the Bogomol'nyi
argument gives a topological lower bound on $E_{\rm cont}$, namely
$$E_{\rm cont} \geq \vert N \vert \eqno(3) $$
(see, for example, [3]). If $\mut=0$, then there are fields which
saturate (3), and which are therefore automatically solutions of the
sigma-model field equations [1,2]. These solutions look like localized
\lq\lq solitons\rq\rq\ in ${\bf R^2}$. Roughly speaking, these solitons are
located at the (isolated) points where $\vp=-k$, ie.~where the field is
antipodal to its boundary value.
But because of the scale invariance of
(2) when $\mut=0$, the size of the solitons is not fixed. So a perturbation
of a soliton can cause it to become arbitrarily concentrated or spread out.
That this indeed happens is confirmed by numerical experiment [9].

One way of stabilizing the soliton is to put in the $\mut$ term, which prevents
spreading out; and also to add to $E_{\rm cont}$ a Skyrme term
$$ E_{\rm Skyrme} = {1\over4\pi}\int_{\bf R^2}
                    \half(\pa_x\vp\wedge\pa_y\vp)^2 \,dx\,dy, \eqno(4) $$
which prevents shrinkage [10,4,11,12,5].
The opposing effects of the $\mut$ and Skyrme terms fix the size of the
soliton (now called a skyrmion)
to be of order $\mut^{-1}$, in the sense that its profile contains a
factor $\exp(-\mut r)$. If $\mut^2=0.1$, then the energy of the $N=1$
skyrmion equals 1.564, i.e. about 50\% higher than the energy of the basic
$N=1$ sigma-model soliton [12,5].
This skyrmion is stable against any perturbation which maintains the
boundary condition and topological charge. Such perturbations can have
arbitrarily large energy, and this represents a very high degree of stability.

Let us return now to the lattice case (1). We impose the boundary condition
that $\vp=k$ on the boundary of the lattice (or $\vp\to k$ if the lattice is
infinite, in which case the condition is necessary to ensure finite energy).
For the time being, put $\mu=0$.

The standard Heisenberg model is obtained by taking $f(\xi)=1-\xi$.
In this case, there is no topology; the configuration space is a connected
manifold. If we remove (by hand) some \lq\lq exceptional\rq\rq\
configurations, then the resulting space does have disconnected components
labelled by an integer $N$, corresponding to the topological sectors of
the continuum case [6]. But if one tries to minimize the energy $E$ within
a nontrivial sector (say $N=1$), then one is driven towards an exceptional
configuration [7]. Another way of putting this is that if one sets up an
initial field which looks like a soliton (ie.~$\vp\approx-k$ at some interior
lattice sites), then it tends to shrink and \lq\lq unwind\rq\rq, and is
certainly not stable.
This feature also appears in numerical simulations of the continuum
sigma model, which in effect involves replacing the continuum by a lattice
[9].

In the continuum, a soliton can be prevented from shrinking by the presence
of the Skyrme term (4). The obvious lattice analogue of this involves
next-to-nearest-neighbour couplings between lattice sites, in addition to
the nearest-neighbour couplings appearing in (1). One model along these lines
was discussed in [8]. This involved adding certain second- and third-nearest
neighbour couplings. With an appropriate tuning of parameters, the lattice
skyrmion then becomes stable (a curious feature is that these more distant
couplings need to be opposite in sign to the nearest-neighbour ones).
But this stability is rather delicate (certainly far weaker than the
topological stability of the continuum skyrmions). A fairly small perturbation
can induce these lattice skyrmions to decay.

The basic reason for the instability, and the absence of topology, is that
there is nothing to prevent neighbouring spins from being wildly different.
In the Heisenberg model it is energetically favourable for a soliton to shrink
to the size of a single lattice cell, in the sense that the spins
$\vp_{x,y}$, $\vp_{x+1,y}$, $\vp_{x,y+1}$, $\vp_{x+1,y+1}$ at the four corners
of this cell
lie on a great circle in the image sphere $S^2$. This is an \lq\lq exceptional
configuration\rq\rq, and from it the field rapidly unwinds.
Clearly in this exceptional configuration there is at least one pair of
neighbouring spins which differ by an angle of at least $\pi/2$.
Conversely, if we could ensure that the angle between neighbouring spins
was always acute, then unwinding could not occur, and topology would be
restored. Such a model is the subject of this paper, and will now be
described.

The idea is to take $f(\xi)$ to be a function which tends to infinity as
$\xi\to0$, for example
$$ f(\xi) = -\log(\xi). \eqno(5) $$
Then if we start with an initial configuration in which the angle between
neighbouring spins is less than $\pi/2$, the field cannot evolve into one which
violates this acuteness condition. If the angle between a pair of neighbouring
spins is $\theta$, then the force which tries to align the spins (ie.~reduce
$\theta$) is proportional to $\tan\theta$; so $\theta$ can never reach $\pi/2$.
By contrast, the analogous force in the Heisenberg model goes like
$\sin\theta$.

In the condensed-matter context, the choice (5) is unphysical. A more physical
choice would be one which increases the energy penalty on pairs of spins
which deviate too much, without involving infinite forces. For example,
one could take
$$ f(\xi) = (1-\xi) + c(1-\xi)^2, \eqno(6) $$
where $c$ is a large positive constant. Such models have indeed been studied
in condensed-matter physics. But they are not topological in the sense of
this paper, and so we shall use (5) in what follows. However, one might expect
that (5) and (6) give solitons with similar properties (under appropriate
conditions). Of course, they have the same continuum limit, namely the
sigma model (2).

The configuration space now consists of all fields $\vp$ such that the angle
between any pair of neighbouring spins is less than $\pi/2$. The energy $E$
is given by (1), with (5), and is a well-defined positive-definite function.
The configuration space is disconnected, with components labelled by an
integer $N$; an algorithm for computing $N$ may be found in [6]. If we start
in the $N$th topological sector and allow the field to evolve, then it has
to remain in this sector: there are infinite potential barriers between the
sectors. And in any given sector there must exist one or more stable solitons,
ie.~fields which minimize $E$.

\bn{\bf3. Numerical Results on Lattice Solitons.}
\sn Our task in this section is to find the minimum-energy configurations
of charges $N=1$ and $N=2$. This involves minimizing a function of
$2(n_x-2)(n_y-2)$ variables, and has to be done numerically. The minimization
is done in two steps. The first step is to assume a \lq\lq radial
symmetry\rq\rq\ with exponentially-decaying profile $g(r)=\pi\exp(-br)$,
and to find the value of $b$ which minimizes the energy $E$ (more details
are given below). In other words, we reduce to a function of the single
variable $b$ and minimize over that. In fact, this gives a field
which is quite close to the true minimum. The second step is to take this
approximation as a starting-point, and minimize over all the degrees of
freedom, obtaining the minimum-energy field to the desired level of
accuracy. This is done by using a conjugate gradient method.

The details of the radial approximation are as follows. The quantity
$$ r = \sqrt{(x-\half n_x-\half)^2 + (y-\half n_y-\half)^2} $$
measures the distance from the centre of the lattice to the site $(x,y)$.
The 3-vector $\vp$ is taken to be
$$ \vp = \bigl(r^{-1}(x-\half n_x-\half)\sin g(r),
               r^{-1}(y-\half n_y-\half)\sin g(r), \cos g(r)\bigr),\eqno(7)$$
except on the boundary of the lattice where it equals $k=(0,0,1)$.
Substituting (7) into (1) gives $E$ as a function of $b$, and this is then
minimized.

Let us first deal with (and dispose of) the case $\mu=0$. On the continuous
plane ${\bf R^2}$, the $\mu$-term is necessary to prevent the skyrmion from
spreading out indefinitely; and the same is true on the infinite lattice.
To be more precise: if we take a square lattice with $n_x=n_y=n$, then the
size of the $N=1$ skyrmion grows like $n/3$ for large $n$. Here
\lq\lq size\rq\rq\ means twice the radius; and the radius is
taken to be the distance
from the centre at which the third component of $\vp$ reaches the value of
0.95 (it equals 1 at the boundary). So the skyrmion is only prevented from
expanding indefinitely by the boundary of the lattice. Conversely, if we want
a skyrmion whose behaviour is not affected by the boundary, then we have to
stabilize its size by taking $\mu$ to be positive. From now on, this is what
we shall do.

By analogy with the continuum case, one might expect the size of a skyrmion
(in the absence of boundary effects) to be proportional to $\mu^{-1}$.
This is indeed the case. One finds numerically that the size grows like
${2\over3}\mu^{-1}$, for $\mu^{-1}$ large (compared to unity). From now on,
$\mu^{-1}$ is set equal to 2. Then the size of the 1-skyrmion is 7
lattice-units; and as long as $n_x$ and $n_y$ are greater than 20, boundary
effects are negligible. To be more precise, changing the size of the lattice
affects the value of the (minimized) energy by less than one part in $10^5$,
and this is the level of accuracy in what follows.

Taking $n_x=n_y=22$, we get a minimum energy of $E=1.57985$. The corresponding
field is depicted in figure 1. What is plotted there (and in all the other
figures) is energy density as a function of $x$ and $y$. So the horizontal
grid represents the lattice, and and we are plotting the quantity
$$\eqalign{
 &-{1\over8\pi}\log(\vp_{x,y}\cdot\vp_{x+1,y})
  -{1\over8\pi}\log(\vp_{x,y}\cdot\vp_{x-1,y}) \cr
 &\quad  -{1\over8\pi}\log(\vp_{x,y}\cdot\vp_{x,y+1})
  -{1\over8\pi}\log(\vp_{x,y}\cdot\vp_{x,y-1})
  +{\mu^2\over4\pi}(1-\vp_{x,y}\cdot k) \cr
}\eqno(8) $$
as a function of $(x,y)$. Summing (8) over all lattice sites gives the total
energy $E$.

Notice that the skyrmion in figure 1 is located at the centre of a lattice
cell, which is also the centre of the whole lattice. If either $n_x$ or
$n_y$ is odd, then in addition to the energy minimum, there also exists
an unstable critical point of $E$. For example, if $n_x=n_y=21$, then
there is a skyrmion \lq\lq balanced\rq\rq\ on the lattice site at the centre
(see figure 2). This has energy $E=1.59888$, ie.~1\% higher than the
minimum. If $n_x=21$ and $n_y=22$, then there is an unstable skyrmion
located at the middle of a link (figure 3). Its energy is $E=1.59033$,
which is 0.7\% higher than the minimum. These energy differences
represent the \lq\lq Peierls-Nabarro potential\rq\rq. If we want to make
the skyrmion move through the lattice, it has to have enough kinetic
energy to overcome the potential barrier between adjacent lattice cells.
Generally speaking, the skyrmion will lose energy as it moves, and eventually
it gets pinned in one position. But this process may have a much longer
time-scale than the dynamics one is investigating.

Finally, let us turn to the $N=2$ case. If $n_x=n_y=22$, the energy is
minimized by the 2-skyrmion depicted in figure 4. It has a broader profile
than the 1-skyrmion, but is still highly localized. Its energy is
$E=2.89439$, which is 8\% lower than twice the energy of the 1-skyrmion.
As before, there are also unstable critical points. If $n_x=n_y=23$,
we get an unstable 2-skyrmion centred at a lattice site (figure 5), with
energy 2.92655 (again 1\% above the minimum). Notice that the energy density
is peaked not at the centre, but on a ring surrounding the centre; this is
a well-known phenomenon in the continuum theory.

\bn{\bf4. Concluding Remarks.}
\sn One application of the idea described above is that it may provide a
highly efficient and economical way of modelling the continuum system.
One should now study the dynamics of these lattice skyrmions, and in
particular compare with numerical simulations of the continuum case
[4,5]. There is of course a choice of dynamics: for example, it could be either
dissipative or non-dissipative.

It is clear that this method extends to the actual Skyrme model, an O(4)
sigma model in three spatial dimensions. Again, the Skyrme term will not
appear explicitly in this lattice theory: its role will be
taken by an appropriate
choice of $f$. The details of this have yet to be investigated.

Is there a lattice version of the Bogomol'nyi bound (3)? For topological
models in one spatial dimension, there are lattice versions where the
lower bound (3) exists and is saturated by static soliton solutions [13].
These have certain advantages, not least that the Peierls-Nabarro barrier
is completely eliminated, and so a soliton can move freely through the lattice.
But in higher dimensions, it seems to be much more difficult to find a
lattice version of (3). So this remains an open question.

\bn{\bf Acknowledgements.}
\sn The author is grateful to H.-R. Trebin and A. Bray for useful
conversations, and to the Isaac Newton Institute for their efficient
hospitality.

\bn{\bf References.}
\item{1.} Belavin, A. A. and Polyakov, A. M., {\sl JETP Letters} {\bf22},
          245 (1975).
\item{2.} Woo, G., {\sl J. Math. Phys.} {\bf18}, 1264 (1977).
\item{3.} Zakrzewski, W. J., {\sl Low dimensional sigma models}
          (Adam Hilger, Bristol, 1989).
\item{4.} Peyrard, M., Piette, B. and Zakrzewski, W. J., {\sl Nonlinearity}
          {\bf5}, 563 \& 585 (1992).
\item{5.} Piette, B. M. A. G., Schroers, B. J. and Zakrzewski, W. J.,
          {\sl Multisolitons in a two-dimensional Skyrme model},
          {\sl Dynamics of baby Skyrmions} (preprints, Durham, 1994).
\item{6.} Berg, B. and L\"uscher. M., {\sl Nucl. Phys. B} {\bf190}, 412 (1981).
\item{7.} L\"uscher, M., {\sl Nucl. Phys. B} {\bf200}, 61 (1982).
\item{8.} Iwasaki, Y. and Yoshi\'e, T., {\sl Phys. Lett. B} {\bf125}, 197
          (1983).
\item{9.} Leese, R. A., Peyrard, M. and Zakrzewski, W. J., {\sl Nonlinearity}
          {\bf3}, 387 (1990).
\item{10.} Bogolubskaya, A. A. and Bogolubsky, I. L., {\sl Phys. Lett. A}
           {\bf136}, 485 (1989).
\item{11.} Izquierdo, J. M., Rashid, M. S., Piette, B. and Zakrzewski, W. J.,
           {\sl Z. Phys. C} {\bf53}, 177 (1992).
\item{12.} Piette, B. M. A. G., M\"uller-Kirsten, H. J. W., Tchrakian, D. H.
           and Zakrzewski,~W.~J., {\sl Phys. Lett. B} {\bf320}, 294 (1994).
\item{13.} Speight, J. M. and Ward, R. S., {\sl Nonlinearity} {\bf7},
           475 (1994).
\bye